\documentclass[11pt,twoside]{article}

\usepackage{asp2004b}
\usepackage{natbib}
\usepackage{epsf}
\usepackage{psfig}
\usepackage{lscape}
\usepackage{graphicx}

\def\5{\footnotesize V\normalsize}
\def\4{\footnotesize IV\normalsize}
\def\3{\footnotesize III\normalsize}
\def\2{\footnotesize II\normalsize}
\def\1{\footnotesize I\normalsize}
\def\lam{$\lambda$}

\begin{document}
\title{VLT-FLAMES observations of young stellar clusters in the Magellanic Clouds} 
\author{C. J. Evans, D. J. Lennon}  
\affil{Isaac Newton Group of Telescopes, Apartado de Correos 321, 
E-38700 Santa Cruz de la Palma, Canary Islands, Spain}
\author{S. J. Smartt}
\affil{The Department of Pure and Applied Physics, The Queen's University of Belfast, Belfast, 
BT7 1NN, Northern Ireland, UK}

\begin{abstract}
We introduce our VLT-FLAMES survey of massive stars in the Galaxy and
the Magellanic Clouds, giving details of the observations in our
younger fields in the LMC and SMC.  In particular we highlight a new
O2.5-type star discovered in N11, and Be-type stars in NGC\,346 with
permitted Fe~\2 emission lines in their spectra.  We give an overview
of the distribution of spectral types in these fields and summarize
the observed binary fraction.
\end{abstract}

\section{Introducing the VLT-FLAMES survey of massive stars}

We have used the Fibre Large Array Multi-Element Spectrograph (FLAMES)
at the Very Large Telescope (VLT) to observe massive stars in 7 fields
in the Galaxy and Magellanic Clouds.  Our target fields were centered
on stellar clusters and were selected to sample a range of metallicity
($Z$) and age.  In practice this means observing fields in the three
most readily accessible, yet differing environments, i.e. the Milky
Way and the Large and Small Magellanic Clouds (LMC and SMC
respectively), as summarized in Table \ref{fields}.

\begin{table}[h]
\caption{Summary of fields observed with VLT-FLAMES\label{fields}}
\smallskip
\begin{center}
\begin{tabular}{lll}
\tableline
\noalign{\smallskip}
& `Young clusters' & `Old clusters' \\
& ($<$5 Myrs)      & (10-20 Myr)    \\
\noalign{\smallskip}
\tableline
\noalign{\smallskip}
Milky Way & NGC 6611 & NGC 3293 \& 4755 \\
\noalign{\smallskip}
LMC       & N11 (inc. LH9/10) & NGC 2004 \\
\noalign{\smallskip}
SMC       & NGC 346 & NGC 330 \\
\noalign{\smallskip}
\tableline
\end{tabular}
\end{center}
\end{table}

FLAMES has 132 science fibres which are fed through to the Giraffe
spectrograph.  Notionally blue targets were selected from pre-imaging
and the FLAMES-Giraffe combination was used to observe a total of 750
stars at high-resolution ($R~\sim$20,000), at six wavelength settings.
These observations form an unprecendented high-quality spectroscopic
survey of OB-type stars, with coverage in the blue optical region from
3850-4750~\AA, and from 6400-6650~\AA, to include the H$\alpha$ Balmer
line.

The primary motivations and observational details of the survey have
been presented by Evans et al. (2005), together with an overview of
the Galactic data.  A number of parallel studies are now underway
using both the Galactic and Magellanic Cloud data to address issues
such as:

\begin{itemize}
\item{To determine the vsin$i$ distributions of the observed stars, is there
evidence for a dependence with $Z$? (cf. Maeder et al. 1999; Keller 2004).}

\item{To obtain physical parameters (including CNO abundances) for a large, 
homogenous sample of OB stars.  These will then be coupled with the vsin$i$ results to
enable comparisons with evolutionary models that attempt to include
the effects of rotation e.g. Maeder \& Meynet (2000, 2001).}

\item{To explore the $Z$-dependence of stellar wind mass-loss rates
in O-type stars (cf. Vink et al. 2001; see de Koter et al., these
proceedings).}

\end{itemize}

\section{Census of N11 and NGC\,346 fields}

Here we focus on the two younger fields in the Magellanic Clouds, N11
and NGC\,346.  These are the richest in the survey in terms of the
numbers of O-type stars, and contain a number of interesting objects.
The distribution of spectral types in these two fields is given in
Table \ref{types}.  

\begin{table}[h]
\caption{Summary of numbers of stars observed in the N11 and NGC\,346 fields\label{types}}
\smallskip
\begin{center}
\begin{tabular}{p{2cm}p{1cm}p{1cm}p{1cm}p{1cm}p{1cm}p{1cm}}
\tableline
\noalign{\smallskip}
Field & O & B0-3 & B5-9 & Be & AFG & Total \\
\tableline
\noalign{\smallskip}
N11     & 43 &   67 &  $-$ &  10 &   4 &   124 \\
\noalign{\smallskip}
NGC 346 & 19 &   59 &    2 &  25 &  11 &   116 \\
\noalign{\smallskip}
\tableline
\end{tabular}
\end{center}
\end{table}

\subsection{A new O2.5-type star in the N11 field}
The N11 field includes the OB associations LH9 and LH10 (Lucke \&
Hodge, 1970), of particular interest in the context of sequential
star-formation.  Prior to our survey the best source of spectroscopic
information in this region was the study by Parker et al. (1992),
which found an apparently younger population in LH10, suggesting that
star-formation in that association was triggered by the evolution of
the most massive stars in LH9.  Owing to observational constraints
(such as crowding) we have not observed all of the stars in the Parker
et al. study, but we have also explored the spectral content in other
nearby regions.

\begin{figure}[h]
\begin{center}
\includegraphics{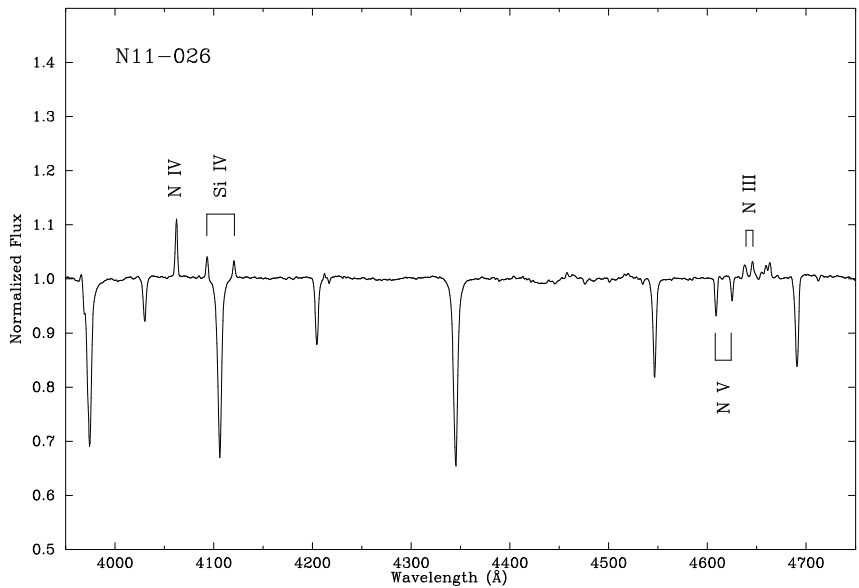}
\caption{A newly discovered O2.5-type star in the LMC; 
for clarity the spectrum has been smoothed by a 1.5 \AA~{\sc fwhm} filter.
The lines identified from left to right are N IV \lam4058; Si IV
\lam\lam4089-4116; NV \lam\lam4604-4620; NIII \lam\lam4634-4640-4642.
\label{o2star}}
\end{center}
\end{figure}

Three O3-type stars were reported in LH10 by Parker et al., all of
which were considered by Walborn et al. (2002) in their extension of
the MK classification scheme to include a new O2 subtype (with one star
from Parker et al. reclassified as an O2 giant).  In principle
the hottest `normal' stars, there are still only $\sim$10 O2-type
stars known.  Here we report the discovery of a new star classified as
O2.5 III(f$^\ast$).  The spectrum (shown in Figure~\ref{o2star}) lies
between the standards published by Walborn et al. (2002, 2004) and so
an intermediate O2.5 subtype is employed.

Interestingly the star is not in LH10, nor in any of the other denser
gas regions in the field, and is $\sim$4.5$'$ to the north of
LH10.  The pre-FLAMES imaging reveals an apparent ionization front in
the nearby gas; moving beyond the central region of LH9 and LH10, there is
still clearly a lot to learn about the star-formation process in this
region.

\subsection{Analysis of OB-type stars}
Of the 62 O-type stars observed in these two fields, half were unknown
prior to this survey; these data form the core of the analysis by
Mokiem et al. (in preparation), which employs an automated approach to
the problem of analysing such a large sample with contemporary model
atmosphere codes (see Mokiem et al., 2005).

The FLAMES survey has also yielded a large number of early B-type
spectra, spanning a range of luminosity classes.  Our final aim is the
consistent analysis of the whole sample; as a first step toward this,
15 of the narrow-lined (i.e. low vsin$i$) stars in NGC\,346, and 21 in
N11, have been analysed by Hunter et al. (in preparation) using the
TLUSTY model atmosphere code (Hubeny \& Lanz, 1995).

\subsection{Be-type spectra}
\begin{figure}[h]
\begin{center}
\includegraphics{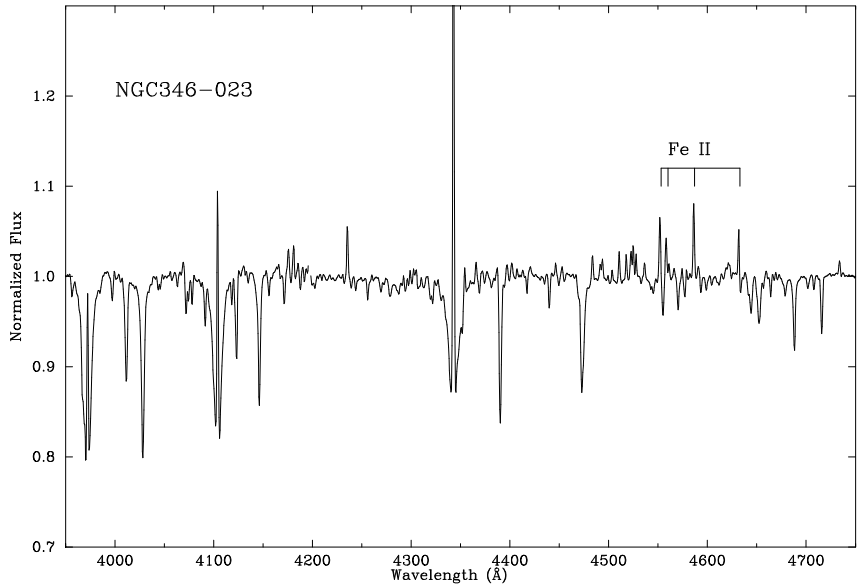}
\caption{A Be-type star in the NGC\,346 field that displays
single-peaked Fe~\2 emission lines; for clarity the spectrum has been
smoothed by a 1.5 \AA~{\sc fwhm} filter.\label{bestar}}
\end{center}
\end{figure}

The observations in NGC\,346 have revealed a number of Be-type stars
with (permitted) emission lines from Fe~\2 in their spectra.  The
morphology of the H$\alpha$ and Fe~\2 profiles suggests that we are
sampling a number of different projection angles; some display
single-peaked emission, whereas others show twin-peaked emission,
commonly interpreted as viewing a circumstellar disk edge-on.
Figure~\ref{bestar} shows the blue-region spectrum of NGC346-023, with
some of the stronger Fe~\2 emission lines marked.  This star is only
2.8$'$ from the centre of the cluster and appears to be a genuine member; 
in a young field such as NGC\,346 one could speculate that these stars
might be somewhat different to classical Be stars, more often associated
with later products of stellar evolution.

With regard to the relative numbers of Be-type stars in
Table~\ref{types}, one should not necessarily conclude that these
provide further evidence for a $Z$-dependence of the Be-type
phenomenon (cf. Maeder et al., 1999); the selection effects for the
Be-stars are not well defined at the current time and we are looking
further into this aspect of the survey.

\section{Binary detection}

\begin{figure}
\begin{center}
\includegraphics{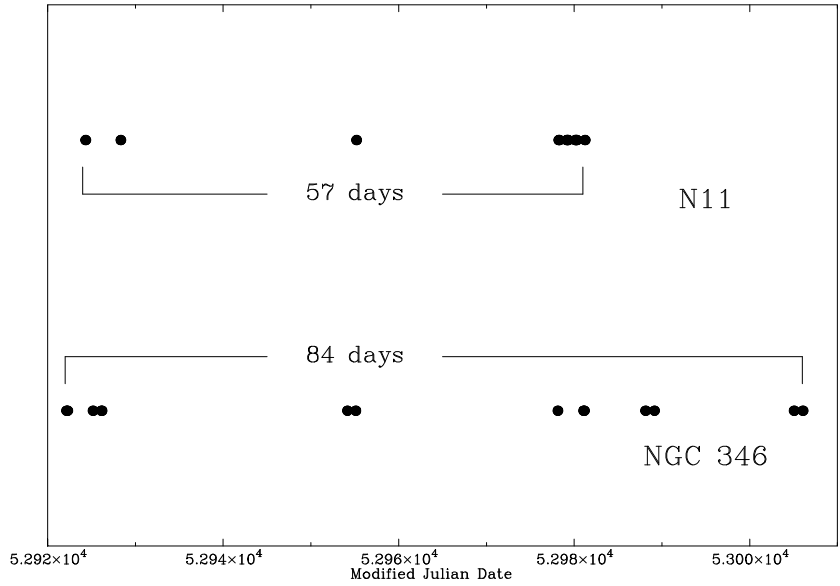}
\caption{Schematic showing the time sampling of the FLAMES observations
in the N11 and NGC\,346 fields.\label{times}}
\end{center}
\end{figure}

Due to the significant scale of the FLAMES survey (in excess of 100 h
VLT time), all of the observations were undertaken in service mode.
In addition to the separate observations at different wavelengths,
each wavelength setting in the LMC/SMC fields was observed at least 6
times (to yield a greater signal-to-noise ratio).  We are therefore
very sensitive to detection of binaries.  Indeed, the sampling is
relatively good in the sense that we have observations at both small
and large time-intervals (as shown in Figure \ref{times}), with the
observations in N11 spanning a total of 57 days and those in NGC 346
covering 84 days.

To date, single and double-lined binaries have been found from
relatively simple methods, i.e. from visual inspection when
classifying the spectra, or from consideration of manual measurements
of the stellar radial velocities from the principal hydrogen, helium
and silicon absorption lines.  More sophisticated methods are now
being used to cross-correlate the individual spectra for each star,
which will likely reveal further binaries (although not expected to be
that significant an increase on the current number).

The FLAMES data are adequate to derive the periods of some of the
newly discovered systems, the more interesting of which will be the
subject of a future study.  Using the numbers in Table \ref{bins}, the
binary fractions in the observed samples are 37$\%$ (N11) and 26$\%$
(NGC\,346).  These data place firm $lower$ limits on the binary
fraction.  The fraction will likely increase slightly following use of
the cross-correlation methods, and naturally some binaries may remain
undetected as a consequence of the time sampling of our observations.

\begin{table}
\caption{Preliminary number of binaries detected in the FLAMES survey,
compared to the total number of stars observed (for stars
classified as earlier than B5, or as Be-type).
\label{bins}}
\smallskip
\begin{center}
\begin{tabular}{p{2cm}p{1cm}p{1cm}p{1cm}p{1cm}}
\tableline
\noalign{\smallskip}
Field   &  O & B0-3 &  Be & Total \\
\tableline
\noalign{\smallskip}
N11     & 43 &   67 &  10 &   120 \\
{\it Binary} & {\it 20} & {\it 20} & {\it 4} & {\it 44}\\
\noalign{\smallskip}
NGC 346 & 19 &   59 &  25 &   103 \\
{\it Binary} & {\it 4} & {\it 18} & {\it 5} & {\it 27}\\
\noalign{\smallskip}
\tableline
\end{tabular}
\end{center}
\end{table}

\section{Summary}

We have given an overview of the spectral content of our FLAMES fields
in N11 and NGC\,346.  The data from these fields alone will form the
basis for numerous studies in the future; indeed parallel analyses are
now underway by various groups on different subsets of the data.  To
underpin these studies a comprehensive catalogue of 
the LMC and SMC fields, including spectral classifications, radial
velocities and detection of binaries will be given by Evans et al. (in
preparation).

\acknowledgements We thank Nolan Walborn for discussions regarding N11-026.
CJE acknowledges support from the UK Particle Physics and Astronomy
Research Council (PPARC) under grant PPA/G/S/2001/00131.  This survey
is based on observations at the European Southern Observatory in
programme 171.0237.

\end{document}